\documentstyle{article}
\begin{document}
\openup6pt 

\title {On static spherically symmetric solutions of the vacuum Brans-Dicke theory} 
\author{Arunava Bhadra \thanks {Email address: aru\_bhadra@yahoo.com} \footnote  {Administrative Block and IUCAA Reference Centre, 
University of North Bengal, Siliguri 734430 
INDIA}  \\
and Kabita Sarkar  \footnote {Department of Mathematics, 
University of North Bengal, Siliguri 734430 
INDIA} \\
}
\date{}
\maketitle

\begin{abstract}
It is shown that among the four classes of the static spherically symmetric solution of the vacuum Brans-Dicke theory of gravity only two are really independent. Further by matching exterior and interior (due to physically reasonable spherically symmetric matter source) scalar fields it is found that only Brans class I solution with certain restriction on solution parameters may represent exterior metric for a nonsingular massive object. The physical viability of the black hole nature of the solution is investigated. It is concluded that no physical black hole solution different from the Schwarzschild black hole is available in the Brans-Dicke theory.  
\end{abstract}


\pagebreak

\section{Introduction} 
Although general relativity (GR) is one of the most beautiful physical theory and is supported by observational evidences, the sustained inability of reconciling GR with quantum mechanics and recent cosmological observations indicate that Einstein's theory needs modification. The Brans-Dicke (BD) theory [1], which describes gravitation through a spacetime metric ($g_{\mu\nu}$) and a massless scalar field ($\varphi$), is a modification of the GR. The theory has recently received interest as it arises naturally as the low energy limit of many theories of quantum gravity such as the supersymmetric string theory or the Kaluza-Klein theory and is also found consistent with present cosmological observations [2]. The theory contains an adjustable parameter $\omega$ that represents the strength of coupling between scalar field and the matter. For certain values of $\omega$, the BD theory agrees with GR in post-Newtonian limit up to any desired accuracy and hence weak-field observations cannot rule out the BD theory in favor of general relativity. It is thus imperative to study strong-field cases in which these two theories could give different predictions.  \\ 
Due to highly non-linear character of all viable gravitational theories, a desirable pre-requisite for studying strong field situation is to have knowledge of exact explicit solution(s) of the field equations. By Birkhoff's theorem, the static spherically symmetric vacuum solution of Einstein's theory is unique, the Schwarzschild metric. On the other hand Birkhoff's theorem does not hold in the presence of a scalar field, hence several static solutions of the BD theory seems possible even in spherically symmetric vacuum situations. Four forms of static spherically symmetric vacuum solution of the BD theory are available in the literature which are constructed by Brans himself [3]. However, it has been shown in [4] that Brans class III and class IV solutions are not different; under a mere redefinition of the radial variable one of them maps to another. \\
Among the all Brans solutions, class I solution is the most studied one as it is the only one which is permitted for all values of $\omega$. The solution in general gives rise to naked singularity [5] though for some particular choices of the solution parameters it represents a black hole different from Schwarzschild one [6]. The other two classes (II and IV) of Brans solution are valid only for $\omega <-3/2$ [3] which implies non-positive contribution of matter to effective gravitational constant and thus a violation of the weak energy condition. However, this energy condition is gradually loosing its status as a kind of law as in many physical situation it could be violated [7-12]. For instance, classical systems built from scalar field non-minimally coupled to gravity violate all the energy conditions [7]. In quantum systems these violations are even more profound. Casimir effect does suggest existence of negative energy density. So is the squeezed states of light [10]. Similarly negative energy density fields also occur in the context of Hawking evaporation of black holes [11], radiation from moving mirrors [12] and in several other situations [10]. The experimental observation of first two effects (Casimir and squeezed states of light) suggests that the idea of negative energy density have to be taken seriously. Besides Solar system observations do not impose any restriction on the sign of $\omega$ [13]. Oppenheimer-Snyder (gravitational) collapse in BD theory also has not conclusively ruled out negative $\omega$ [14]. Hence class II and IV solutions can not be regarded as unphysical just for being negative $\omega$ solutions. These solutions or their Einstein frame variants [15, 4] have been used in literatures in different contexts such as in connection with the wormhole physics [16,17] or to generate solutions of the string theory (in string frame) [18]. In some aspects Class IV solution, which gives rise to what is called as cold black hole [19, 20], even exhibits better behavior than class I solution. For example the tidal forces do not diverge on the horizon for this spacetime unlike the class I metric [20].  \\
All Brans solutions, however,  may not be physically relevant. There are several known exact perfect fluid interior solutions in general relativity [21]. Most of them are physically not acceptable because either the solutions have not a well defined boundary or they do not match with the Schwarzschild exterior solution at the boundary surface or perfect fluid satisfies unrealistic equations of state or due to some other valid reasons. The same could happen for the vacuum BD solutions also. In the present article we would like to examine the physical viability of the Brans solutions. The paper is organized as follows. After giving a short account of the BD theory and its static spherically symmetric solutions, it has been shown in section 2 that Brans class I and class II solutions are essentially not different. Physical viability of Brans solutions has been examined in section 3. The broad nature of the class I solution has been discussed in section 4 and the physical relevance of the Brans class I black hole has been investigated in the section 5. Finally the results are summarized in section 6. \\

\section{Static spherically symmetric vacuum solutions of the BD theory}
In the BD theory the scalar field acts as the source of the (local) gravitational coupling with $G \sim \varphi^{-1}$ and consequently the gravitational {\lq constant \rq} is not in fact a constant but is determined by the total matter in the universe through an auxiliary scalar field equation. The scalar field couples to both matter and spacetime geometry and as mentioned before the strength of the coupling is represented by the dimensionless constant $\omega$. The theory is found consistent with the (local) observations only when $\omega$ is very large. A lower limit $\vert \omega \vert > 5 \times 10^4$ is obtained from the recent conjunction experiment with Cassini spacecraft [22]. This suggests even if scalar field exists, predictions of the BD theory are not much different from GR, particularly in the weak-field regime, because under the limit $ \vert \omega \vert \rightarrow \infty$, the BD theory (and its dynamic generalization) reduces to GR [23] unless the matter field is traceless [24]. \\
In the Jordan conformal frame, the BD action takes the form (we use geometrized units such that $G=c=1$ and we follow the signature -,+,+,+)
\begin{equation}
{\cal A}= \frac{1}{16 \pi }\int d^{4}x \sqrt{-g}\left(\varphi R+\frac{\omega }{\varphi } 
g^{\mu\nu} \varphi_{,\mu} \varphi_{,\nu} + {\cal L}_{matter} \right)
\end{equation}
where \( {\cal L} _{matter} \) is the Lagrangian density of ordinary matter. Variation of (1) with 
respect to $ g^ {\mu \nu} $ and $\varphi$ gives, respectively, the field equations
\begin{equation} 
R_{\mu\nu} -\frac{1}{2}g_{\mu\nu}R= -\frac{8 \pi} {\varphi } T_{\mu \nu}
-\frac {\omega}{\varphi ^{2}}\left( \varphi_{,\mu} \varphi_{,\nu}- \frac{1}{2} g_{\mu\nu} 
\varphi^{,\sigma} \varphi_{,\sigma} \right) - 
\frac{1}{\varphi} \left( \nabla_{\mu}\nabla_{\nu}\varphi-g_{\mu \nu} \Box \varphi \right),  
\end{equation}
\begin{equation}
\Box \varphi = \frac {8\pi T}{(2\omega + 3)} 
\end{equation}
where $R$ is the Ricci scalar, and $T$=$T_{\mu}^{\mu}$ is the trace of the matter energy momentum tensor. \\
As stated earlier, Brans provided four classes of static spherically symmetric solutions of the above theory when $T_{\mu \nu}=0$. The Brans class I solution (in isotropic coordinates) is given by 
\begin{equation}
ds^{2}= -e^{\alpha_{o}}\left( \frac{1-B/\rho}{1+B/\rho} \right)^{\frac{2}{\lambda}} dt^{2} + e^{\beta_{o}}\left( 1 + \frac{B}{\rho}\right) ^{4} \left( \frac{1-B/\rho}{1+B/\rho} \right)
^{\frac{2(\lambda -C -1)}{\lambda}} \left( d\rho^{2} +\rho^{2} d\theta ^{2} +\rho^{2} sin^{2} \theta d\phi ^{2} \right)
\end{equation}
\begin{equation}
\varphi = \varphi_{0} \left( \frac{1-B/\rho}{1+B/\rho} \right)^{\frac{C}{\lambda}}
\end{equation}
with the constraint condition
\begin{equation}
\lambda^2 =  (C+1)^{2} - C(1-\frac{\omega C}{2})
\end{equation}
where $\alpha_{o}, \beta_{o}, B, C $ are arbitrary constants. \\
The class II solution is given by
\begin{equation}
ds^{2}= -e^{\alpha_{o} + \frac{4}{\Lambda} tan^{-1}(\rho^{'}/B^{'}) } dt^{2} + e^{\zeta_{o} - \frac{4(C+1)}{\Lambda} tan^{-1}(\rho^{'}/B^{'}) -2ln[\rho^{'2}/(\rho^{'2}+B^{'2})]} \left( d\rho^{'2} +\rho^{'2} d\theta ^{2} +\rho^{'2} sin^{2} \theta d\varphi ^{2} \right)
\end{equation}
\begin{equation}
\varphi = \varphi_{0}e^{2C/\Lambda tan ^{-1}(\rho^{'}/B^{'})} 
\end{equation}
with the solution parameters are related by
\begin{equation}
\Lambda^2 =   C(1-\frac{\omega C}{2}) -(C+1)^{2} 
\end{equation}
Here arbitrary constants are denoted as $\alpha_{o}, \zeta_{o}, B^{'}, C $. \\
Class III solution can be written as
\begin{equation}
ds^{2}= -e^{\alpha_{o}-2\rho /B} dt^{2} + e^{\beta_{o} -4ln(\rho /B + 2(C+1)\rho /B}\left( d\rho^{2} +\rho^{2} d\theta ^{2} +\rho^{2} sin^{2} \theta d\phi ^{2} \right)
\end{equation}
\begin{equation}
\varphi = \varphi_{0} e^{C\rho/B}
\end{equation}
with the condition
\begin{equation}
C= \frac{-1 \pm \sqrt{-2\omega -3}}{\omega +2}
\end{equation}
and finally class IV solution is
\begin{equation}
ds^{2}= -e^{\alpha_{o}-2/(B\rho) } dt^{2} + e^{\beta_{o} + 2(C+1)/(B\rho ) }\left( d\rho^{2} +\rho^{2} d\theta ^{2} +\rho^{2} sin^{2} \theta d\phi ^{2} \right)
\end{equation}
\begin{equation}
\varphi = \varphi_{0} e^{C/(B\rho)}
\end{equation}
with
\begin{equation}
C= \frac{-1 \pm \sqrt{-2\omega -3}}{\omega +2}
\end{equation}
The class I and class II solutions are, however, not different. To show this we define a new radial variable $\rho=1/\rho^{'}$. Utilizing the identity $tan^{-1}(x)=\frac{i}{2}ln \left( \frac{1-ix}{1+ix} \right)$, the Eqs.(7) and (8) can be recast as 
\begin{equation}
ds^{2}= -e^{\alpha_{o}} \left( \frac{1-iB^{''}/\rho}{1+iB^{''}/\rho} \right)^{\frac{2i}{\Lambda}} dt^{2} + e^{\beta_{o}}\left( 1 + \frac{B^{''2}}{\rho^2}\right) ^{2} \left( \frac{1-iB^{''}/\rho}{1+iB^{''}/\rho} \right)
^{\frac{-2i(C+1)}{\Lambda}} \left( d\rho^{2} +\rho^{2} d\theta ^{2} +\rho^{2} sin^{2} \theta d\phi ^{2} \right)
\end{equation}
\begin{equation}
\varphi = \varphi_{0} \left( \frac{1-iB^{''}/\rho}{1+B^{''}/\rho} \right)^{\frac{iC}{\Lambda}}
\end{equation}
where $B^{''}=1/B^{'}$, and $\beta_{o}=\zeta_{o}-4lnB^{''}$. The above two equations would reduce to Eqs. (4) and (5) if we denote $\lambda=-i \Lambda$ and $B=iB^{''}$. The relation (9) will also map to relation (6) under such redefinition of constants. Therefore, class I and class II solutions are equivalent; choice of imaginary $B$ and $\lambda$ in class I solution leads to the class II solution. A point to be noted is that for imaginary $B$ and $\lambda$ the class I solution becomes regular at all points including the point $r=B$ and consequently the class II solution does not possess any horizon.  \\
As mentioned before Brans class III and class IV solutions are also not different [4]; under a mere redefinition of the radial variable ($\rho \equiv 1/\rho$) one of them maps to another. Hence only two classes of solutions, class I and class IV, are found independent. These two classes can be expressed by a single form. For this purpose let us consider the transformation 
\begin{equation}
e^{-\sigma/r}=\frac{1-B/\rho}{1+B/\rho}
\end{equation}
($\sigma \ne 0$) under which the Eqs.(4) and (5) are reduced to the form
\begin{equation}
ds^{2}= -e^{\alpha_{o}-\alpha/r}dt^{2} + e^{\zeta_{o}+\alpha (C+1)/r}\left( \frac{\sigma/r}{sinh (\sigma/r)}\right) ^{4}dr^{2} + e^{\zeta_{o} + \alpha (C+1)/r}\left( \frac{\sigma/r}{sinh (\sigma/r)}\right) ^{2} r^{2} \left( d\theta ^{2} + sin^{2} \theta d\phi ^{2} \right)
\end{equation}
and
\begin{equation}
\varphi = \varphi_{0}e^{-\frac{\alpha C}{2r}} 
\end{equation}
$ \zeta_{o}= \beta_{o} +ln(B^4/ \sigma^2) $. The relation among the parameters is given by
\begin{equation}
4\frac{\sigma^2}{\alpha^2} =  (C+1)^{2} - C(1-\frac{\omega C}{2})
\end{equation}
The Eqs. (19)-(21) are the general form of all Brans' solutions. As is evident from the above $\sigma \ne 0$ leads to class I solution where use of some imaginary parameters results the class II solution. The choice $\sigma=0$ gives the Brans' class IV solution and a further redefinition of the radial variable $\bar{r}=1/r$ will immediately give the class III solution. The line element (19) is conformal to the Wyman solution [25] of the Einstein minimally coupled scalar field theory. \\

\section{Physical viability of Brans solutions}
In general relativity the metric tensor is the only gravitational field variable. Hence, it is sufficient and necessary to match the interior and exterior solutions for metric tensor only. In contrast BD theory has additional scalar field which contributed to the gravitational field as well. Therefore, in this theory not only matching for metric tensor is necessary but also for the additional scalar field. \\
It follows from Eq. (3) that to the leading order in $1/r$ the interior (in presence of matter) scalar field satisfies the equation
\begin{equation}
\nabla^{2} \stackrel{2}{\varphi} =  -\frac{8\pi}{2\omega+3}\stackrel{o}{T} 
\end{equation}
where we have expanded scalar field as $\varphi = \varphi_{o} + \stackrel{2}{\varphi}+ \stackrel{4}{\varphi}+ ...$,  $\stackrel{N}{\varphi}$ denotes the term in $\varphi$ of order $\frac{1}{r^{N/2}}$ (here we have followed the notation of [23]) and $\stackrel{o}{T}$  denotes the term in $T^{\sigma}_{\sigma}$ of order $\frac{1}{r^{3}}$ ($\stackrel{o}{T}^{oo}$ is the density of rest mass) [23]. Therefore to the leading order in $1/r$ (r being the radial variable) the scalar field at near the surface of the matter distribution has the following expression 
\begin{equation}
\varphi = \varphi_{o} - \left(\frac{2}{2\omega+3} \right) \phi
\end{equation}
where $\phi$ is the Newtonian potential defined through $\nabla^{2} \phi = 4\pi \stackrel{o}{T}$. At the surface and outside the source $\phi = -M/r$ where $M$ is the total mass of the source as viewed by a distant observer. Utilizing the relation $G=\frac{2\omega+4}{2\omega+3}\frac{1}{\varphi_{o}}$ (which can be obtained from the relation $\stackrel{2}{g} _{oo} =  -2 \phi $) where $G$ is the gravitational constant that would be measured in a real experiment, we finally get the expression for scalar field to the leading order in $1/r$ near the surface of the matter distribution (we still continue of using geometrized units)
\begin{equation}
\varphi = \varphi_{o} \left( 1+\frac{1}{\omega+2} \frac{M}{r} \right) 
\end{equation}
A physically viable external solution for scalar field must match smoothly with the above expression at the surface. \\
When matching to two different solutions on a common surface it is essential to choose an appropriate coordinate system. The expression (24) is written in standard coordinates ($t, r, \theta$ and $\phi$). Hence the expression for scalar field in Brans class I or class IV solution needs to transform first in the standard coordinates for effective comparison. However, at the first order the standard and isotropic coordinates produce identical effects. Hence comparing the expression for scalar field of Brans class I solution with Eq. (24) and using the relation $\stackrel{2}{g} _{oo} =  -2 \phi $, we get 
\begin{equation}
C=-\frac{1}{\omega +2}; \; 2B/\lambda=M
\end{equation}
The relation (6) gives $\lambda=\sqrt{ \frac{2\omega +3}{2 \omega +4}} $. Therefore, Brans class I solution  may represent external gravitational field due to a reasonable matter field only when the solution parameters are given by Eq. (25). On the other hand scalar field of class IV solution can not be matched with interior solution at the surface as its 1st order term goes as $\omega^{-1/2}$ for large $\omega$ whereas boundary condition requires that it must be proportional to $\omega^{-1}$. Thus regularity conditions at the boundary suggest that class IV solution cannot be acted as an exterior metric for a nonsingular spherical massive object. \\

\section{Generic nature of Brans class I solution} 
In dealing with scalar-tensor theories in general and BD theory in particular, one envisages two types of frames, viz., the Jordan and Einstein frames which are connected via the conformal relation [26] 
\begin{equation}
\tilde{g}_{\mu \nu}=\varphi g_{\mu \nu}
\end{equation}
and a redefinition of the scalar field 
\begin{equation}
d\tilde{\varphi}=\left[\frac {2\omega+3}{2 \lambda}\right]^{\frac{1}{2}} \frac{d\varphi}{\varphi}
\end{equation}
where $\lambda$ is a constant and $\tilde{g}_{\mu \nu}$ and $\tilde{\varphi}$ are the Einstein frame variables. Though experimentally observed quantities are those that are written in the Jordan frame [27] sometimes it is mathematically more preferable to use the Einstein representation as the spin-2 and spin-0 fields are decoupled in the later frame from each other and the behavior of the fields are more readily manageable. In this conformal frame the field equations are given by
\begin{equation} 
\tilde{R}_{\mu \nu} -\frac{1}{2}\tilde{g}_{\mu \nu}\tilde{R}=-\lambda T_{\mu \nu}, \; \tilde{\varphi}^{;\sigma}_{;\sigma} = 0 ,  
\end{equation}
and
\begin{equation} 
T_{\mu \nu}=\tilde{\varphi}_{,\mu}\tilde{\varphi}_{,\nu}-\frac{1}{2}\tilde{g}_{\mu \nu}\tilde{\varphi}^{,\sigma}\tilde{\varphi}_{,\sigma} ,  
\end{equation}
where $\tilde{R}$ is the Ricci scalar, $T_{\mu \nu}$ is the energy momentum tensor due to the massless scalar field $\tilde{\varphi}$. With the assumption of asymptotic flatness and taking scalar field to be time independent ($\dot{\varphi}=0$), Wyman, by directly solving the field equations in a straightforward way, has shown [25] that the most general static spherically symmetric metric that satisfies the above field equations is given by
\begin{equation}
ds^{2}= -e^{\alpha/r}dt^{2}  +e^{-\alpha /r}\left( \frac{\sigma/r}{sinh (\sigma/r)}\right) ^{4}dr^{2} + e^{- \alpha /r}\left( \frac{\sigma/r}{sinh (\sigma/r)}\right) ^{2}  r^{2} \left( d\theta ^{2} + sin^{2} \theta d\phi ^{2} \right).
\end{equation}
Later Roberts has shown [28] that the assumption of asymptotic flatness is not even required for obtaining the above general solution. Note that the well known static spherically symmetric solution of the Einstein minimally coupled scalar field theory, the Buchdahl solution [29], which is also variously referred [15] to as Janis-Newman-Winicour solution [5], is contained in the Wyman solution as a special case. \\
As mentioned already, the Wyman solution is conformal to the general form of Brans' solutions (Eqs.(19) to (21)). Since the mapping between the two conformal frames, the Jordan and Einstein, is one-to-one, Eqs. (19) -(21) thus should be the general static spherically symmetric solution of the BD theory in the Jordan frame. Here it should be noted that though look very similar, the characteristics of the solutions in the Einstein frame representation are quite different from those in the Jordan frame. For instance, the Buchdahl solution, which is conformal to the Brans class I solution,  always satisfies weak energy condition and exhibits strong globally naked singularity for any choice of the solution parameters unlike the Brans class I solution. As it is already shown that among the Brans solutions only class I solution with parameters constrained by the Eq.(25) represents exterior metric for a nonsingular spherical massive object, hence though the class I metric is not the unique solution of the BD theory but it is the most general physically acceptable static spherically symmetric solution of the theory. In the limit $\omega$ tends to $\infty$ this solution reduces to the Schwarzschild metric with constant scalar field. Recently He and Kim [30] have claimed for two {\it new} static vacuum  solutions of the BD theory. But as already shown in [31, 32], these two classes of solutions are essentially limiting cases of the Brans class I solution. \\

\section{Physical relevance of Brans class I black hole} 
In general, class I solution exhibits naked singularity; all curvature invariants diverge at the horizon $\rho=B$. But for this reason one cannot rule out the solution as whether a naked singularity occurs generically in a physical realistic collapse is still a subject of considerable debate [33]. However, as demonstrated by Campanelli and Lousto [6], class I solution also exhibits black hole nature when solution parameters obey certain constraint conditions as mentioned below. Under such restrictions  the Hawking temperature for the metric becomes zero and hence the solution is recognized as cold black hole following the terminology of Bronnikov {\it et al} [19]. It was further shown in [6] that some strong gravitational fields effects such as scattering of photons, X-ray luminosity of accretion disks, or Hawking radiation could distinguish the BD black holes from the Schwarzschild one. \\
Campanelli and Lousto use the following form of the solution
\begin{equation}
ds^{2}= -\left(1-\frac{B}{r} \right)^{m+1} dt^{2} - \left(1-\frac{B}{r} \right)^{n-1}dr^{2} - r^{2} \left(1-\frac{B}{r} \right)^{n} \left(d\theta ^{2} + sin^{2} \theta d\phi ^{2} \right)
\end{equation}
and
\begin{equation}
\varphi = \varphi_{0}\left(1-\frac{B}{r} \right)^{-(m+n)/2} 
\end{equation}
where m,n are arbitrary constants. The coupling constant is related with the parameters as is given by
\begin{equation}
\omega =-2 \frac{m^2+n^2+nm+m-n}{(m+n)^{2}}
\end{equation}
The equations (31) to (33) transform to the original form of the solution as given by equations (4) to (6) under the radial transformation
\begin{equation}
r=\rho \left(1+\frac{B}{\rho}\right)^{2}
\end{equation}
and with the identification $m=1/\lambda -1$, $n= 1-\frac{C+1}{\lambda}$. Campanelli and Lousto found that for $ n \le -1$, class I solution admits black hole space time. This can be understood by studying the curvature invariants of the metric. The Ricci scalar for the class I metric (4) is given by
\begin{equation}
R=\frac{4 \omega B^2 C^2 }{\lambda^2 \rho^{4} (1+B/\rho)^{8}}  \left( \frac{1-B/\rho}{1+B/\rho} \right)
^{\frac{-2(2\lambda -C -1)}{\lambda}}
\end{equation}
The expression for the Kretschmann scalar is quite messy. To the leading order of 1/r it has the expression
\begin{equation}
R_{\alpha \beta \gamma \delta}R^{\alpha \beta \gamma \delta}\approx \frac{96B^2(2 +2C+ C^2)}{\lambda^2 \rho^{6} (1+B/\rho)^{16}}  \left( \frac{1-B/\rho}{1+B/\rho} \right)
^{\frac{-4(2\lambda -C -1)}{\lambda}}
\end{equation}
and the Weyl scalar is given by
\begin{equation}
C_{\alpha \beta \gamma \delta}C^{\alpha \beta \gamma \delta}= \frac{16 B^2 \left[2B(4+ 4C+ C^2 +2\lambda^{2})r-3\lambda (C+2)(B^{2}+r^{2} ) \right]^2  }{3 \lambda^{4} \rho^{10} (1+B/\rho)^{16}}  \left( \frac{1-B/\rho}{1+B/\rho} \right)
^{\frac{-4(2\lambda -C -1)}{\lambda}}
\end{equation}
It can be seen easily that as $r \rightarrow B$ all curvature invariants diverge and the solution exhibits naked singularity. However, when 
\begin{equation}
C+1 \ge 2 \lambda
\end{equation} 
curvature invariants become non-singular. If further $(C+2-\lambda)/\lambda > 0$ then the surface $\rho=B$ will be an outgoing null surface and hence it will act as event horizon and the solution exhibits black hole nature [6]. Note that the condition (38) implies $\omega$ to be negative and thus a violation of the weak energy condition. Now if we impose the restriction (25) which is needed for class I solution to be physically acceptable, the inequality (38) for real $C$ and $\lambda$ demands $ -2 < \omega < -(2+\frac{1}{\sqrt{3}})$. Such small values of $ \omega $ is already ruled out by observations. 

\section{Discussion}
A viable theory of gravity could have several exact explicit solutions. Though many of those solutions may be useful for understanding the inherent non-linear character of gravitational theories, only physically acceptable solutions are of astrophysical interest. In this work we examine different classes of the static spherically symmetric solution of vacuum BD theory for their physical relevance.\\
It has been found that among  the four different forms of the static spherically symmetric solution of the vacuum BD theory of gravity only two classes, Brans class I and class IV solutions, are really independent; the remaining solutions are their variant. Moreover, by matching the expressions for scalar field of the independent Brans solutions with the interior solution for the scalar field due to physically reasonable matter source, it is found that only Brans Class I solution may represent external gravitational field for nonsingular spherically symmetric matter source when the parameters of the solution have a specific dependence on coupling constant $\omega$ as given by Eq. (25). The class IV solution, though also admits all the standard weak field tests (up to the first post-Newtonian order) of gravitation, does not act as an exterior metric for any reasonable gravitating object.    \\
Hawking theorem [34] states that the static spherically symmetric black hole solution of the BD theory is the Schwarzschild black hole. The proof relies on the assumption that the weak energy condition holds. Campanelli and Lousto demonstrated that the BD theory admits black holes different from the Schwarzschild one when the weak energy condition is not respected. As it is well known now that in many physical situation the weak energy condition could be violated, the existence of BD black holes in nature is an interesting possibility. The present investigation, however, suggests that such black holes are physically not viable; they are incompatible with the observationally imposed constraints on the solution parameters. \\

\noindent {\bf Acknowledgments}
Authors thank an anonymous referee for useful comments and suggestions. \\

\end{document}